\documentclass[]{rsos}%%%%where rsos is the template name

\usepackage{graphicx}
\usepackage{xcolor}
\usepackage{adjustbox}        % rotate stuff
\renewcommand\thesection{\arabic{section}}

\begin{document}
%%%%%%%%%%%%%%%%%%%%%%%%%%%%%%%%%%%%%%%%%
\title{Self induced class stratification in 
competitive societies of agents: Nash stability 
in the presence of envy}

\author{Claudius Gros}
\address{Institute for Theoretical Physics, Goethe University Frankfurt,
Frankfurt a.\,M., Germany}

%%%% Subject entries to be placed here %%%%
\subject{systems theory, mathematical modeling}

%%%% Keyword entries to be placed here %%%%

%%%% Insert corresponding author and its email address}
\corres{Claudius Gros\\
\email{gros07[at]itp.uni-frankfurt.de}}

\begin{abstract} % max 200 words
Envy, the inclination to compare rewards, can be expected to unfold when
inequalities in terms of payoff differences are generated in competitive
societies. It is shown that increasing levels of envy lead inevitably to a
self-induced separation into a lower and an upper class.  Class stratification
is Nash stable and strict, with members of the same class receiving identical
rewards. Upper class agents play exclusively pure strategies, all lower class
agents the same mixed strategy. The fraction of upper class agents decreases
progressively with larger levels of envy, until a single upper class agent is
left. Numerical simulations and a complete analytic treatment of a basic
reference model, the shopping trouble model, are presented. The properties of
the class-stratified society are universal and only indirectly controllable
through the underlying utility function, which implies that class stratified
societies are intrinsically resistant to political control. Implications for
human societies are discussed. It is pointed out that the repercussions of envy
are amplified when societies become increasingly competitive.
\end{abstract}

\begin{fmtext}    % first page

%%%%%%%%%%%%%%%%%%%%%%%%%%%%
\section{Background}
%%%%%%%%%%%%%%%%%%%%%%%%%%%%

Is it possible that societies separate on their own into 
distinct social classes when everybody is otherwise 
interchangeable, born equal? This is the question examined 
here. Being equal means in a game theoretical setting 
that agents have access to the same options and 
payoff functions. Starting with random 
initial policies, strategies evolve according to the 
payoff received on the average. For our investigation 
we assume that three constituent features characterize 
the payoff function. Firstly, options come with a range 
of distinct payoffs.
\end{fmtext}
%%%%%%%%%%%%%%% End of first page %%%%%%%%%%%%%%%%%%%%%
\maketitle

\noindent
Secondly, competition for resources 
is present, which implies that agents selecting the same 
option are 
penalized. Thirdly, players care how they are 
doing with respect to others. We show that an endogenous
transition to a strictly class stratified society takes
place when these three conditions are fulfilled.

It is well established that people live not in isolation,
but that social context influences memory, cognition
and risk taking in general 
\cite{wyer2014memory,linde2012social,lahno2015peer}, 
that it leads to accountability \cite{tetlock1985accountability} 
and to group decision making \cite{tindale2019group}. A key
aspect of social context is the quest for social status
\cite{frank1985choosing,courty2019pure}, which has been 
modeled using several types of status games 
\cite{akerlof1997social,oxoby2003attitudes}.
Of particular relevance to our approach is the notion
that the satisfaction an individual receives from 
having and spending money depends not only on the 
absolute level of consumption, but also on how 
this level compares with that of others 
\cite{hopkins2004running}. This view has seen
widespread support from relative income theory
\cite{mcbride2001relative,clark2010compares}.
Relative gauges are considered similarly to 
be of relevance for the definition of poverty
\cite{sen1983poor,wagle2002rethinking}.

The outcome of a game may be considered fair in
a social context when nobody has an incentive to
trade the reward received. For the problem of 
allocating multiple types of goods, which may
be either divisible or indivisible, like apples, 
banana and kiwis, the outcome is said to be free of 
envy when the recipients are content with their bundles
\cite{foley1967resource,nguyen2014minimizing}. 
Here we use envy in analogy to denote the 
propensity to compare rewards between agents.
When relative success is important it implies
that the payoff function is functionally dependent 
on the outcome, the average payoff received.
A feedback loop is such established. It is well 
known, e.g.\ from the theory of phase transitions 
in physics \cite{stanley1971phase}, that feedback 
loops can lead to collective phenomena and hence
to novel states. Indeed we find that envy induces 
a new state, a self-induced class stratified Nash 
equilibrium. 

We examine here the interaction between social context
and competition for scarce resources, which lies at
the core of many games. A typical example is the 
Hawks and Doves framework, for which the reward is
divided when both agents select the same behavioral
option. In a society of agents a range of options 
yielding distinct payoffs will be in general available. 
In this setting, competition may force agents to 
select different strategies, for instance to settle for 
the second best course of action when the option with the 
highest prospective reward has already been claimed by 
somebody else. The 
outcome is a multi-agent Nash state, forced cooperation, 
in which agents seemingly cooperate by avoiding each 
other, but only because it pays off and not out of sheer
good will. Other forms of cooperation \cite{hauert2005game},
such as reciprocal altruism \cite{trivers1971evolution}
and indirect reciprocity \cite{nowak1998evolution},
share this trait. A key aspect of forced cooperation is 
that it is unfair in terms of reward differentials, the 
precondition for envy to take effect. 

Forced cooperation can be argued to a generic feature of 
real-world societies, both when agents are differentiated 
or not. In ecology, for instance, non-uniform resource allocation 
is observed in competitive population dynamics models when
resources are scarce \cite{anazawa2019inequality}. Envy has
hence the potential to induce novel societal states 
in which just the initial conditions, and not 
differences between agents per se, determine in 
which class someone ends up. In previous 
studies, class structures have been presumed to exist 
\cite{haagsma2018income}, or to be dependent on 
as-of-birth differences \cite{akerlof1997social}.
Clustering into distinct classes may occur
similarly for networks of agents when comparison 
is restricted to neighbors \cite{immorlica2017social}.

Outcome and input, the reward received and the structure 
of the payoff function, are interdependent when envy is 
present, a setup that is typical for dynamical systems 
studied by complex systems theory \cite{gros2015complex}. 
Key aspects of the present investigation, including
in part terminology and analysis methods, are hence 
based on complex systems theory. One could also ask 
if it would be feasible to optimize properties of the 
stationary state considered desirable, such
as fairness, as done within mechanism design theory 
\cite{maskin2008mechanism}. An example would be to 
set incentives for prosocial behavior \cite{benabou2006incentives},
with the overall aim to optimize society 
\cite{myerson1981utilitarianism}. This is a 
highly relevant program. But what if the stationary
state of the society has in part universal properties 
that cannot be altered by changing the underlying 
utility function, being independent of it? We find, 
that this is precisely what happens when envy is relevant.

Our basic model is motivated by a shopping analogy.
A clique of friends gathers for an exclusive wine 
tasting, with everyone shopping beforehand. There
are several wine outlets, each specialized in a 
specific quality. In the wine cellars a wide selection 
of vintage years are kept in storage, but only 
a single bottle per year. Shopping in the same wine 
cellar as somebody else implies then that someone 
has to content with second-best vintage year. At 
the gathering, the friends enjoy the wine, becoming 
envious if somebody else made the better deal.
Both extensive numerical simulation of the shopping
trouble model and an encompassing analytic treatment 
of the class stratified state are presented. An
overview of the terminology used is given in
Sect.~\ref{sect_termiology}.

%%%%%%%%%%%%%%%%%%%%%%%%%%%%%%%%%%
\begin{figure}[t!]
%\centerline{
%\includegraphics[width=1.00\columnwidth,angle=0]{fig_100_100_03_00_example.pdf}
%           }
%\centerline{
%\includegraphics[width=1.00\columnwidth,angle=0]{fig_100_100_03_00.pdf}
%           }
\centerline{
\includegraphics[width=1.00\columnwidth,angle=0]{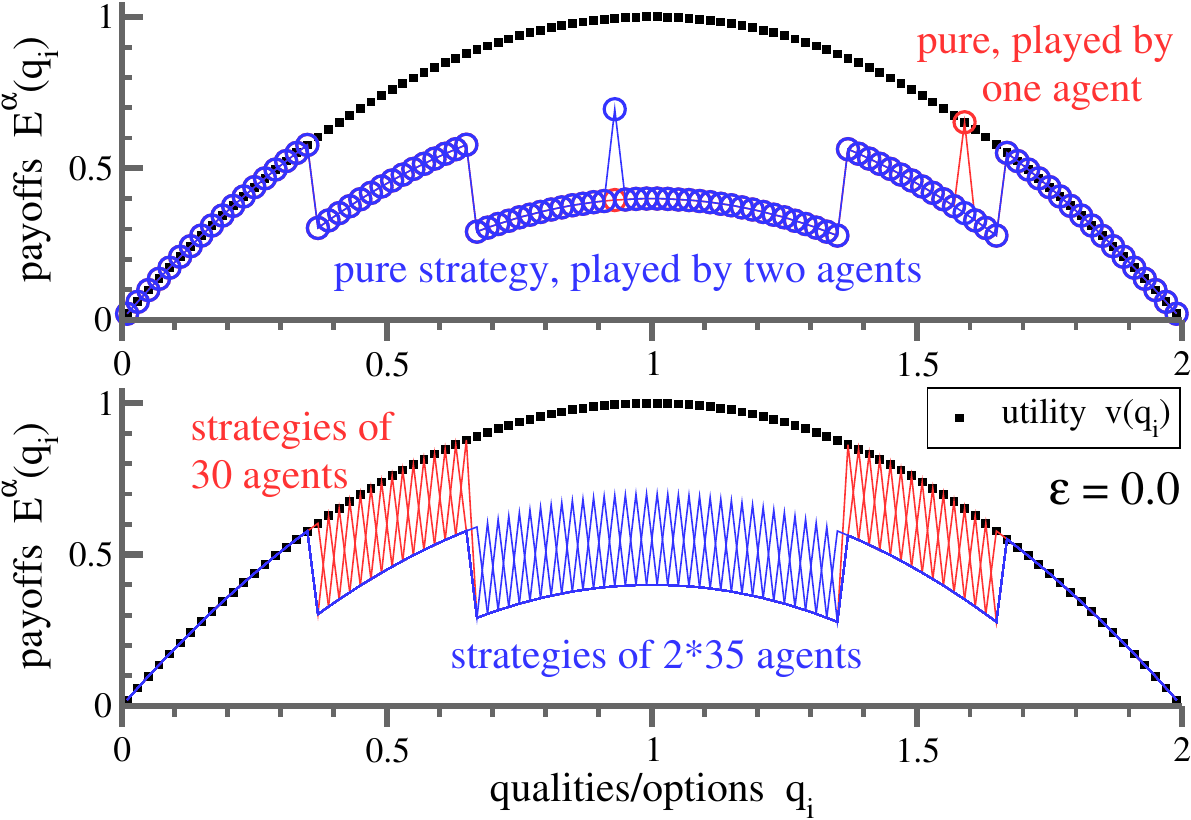}
           }
\caption{{\bf Forced cooperation.} The shopping trouble
model in the absence of envy, $\varepsilon\!=\!0$. For
$\kappa\!=\!0.3$, $M\!=\!100$ agents and $N\!=\!100$ 
options, the payoffs $E^\alpha(q_i)$ (connected by lines) 
obtained using evolutionary dynamics are shown.
The underlying utility function, $v(q_i)=1-(1-q_i)^2$,
has been included as a reference (black squares). Only
pure strategies are played, which implies that 
the rewards $R^\alpha$ correspond to the peaks of the 
respective payoffs, compare (\ref{R_alpha}). A total 
of 65 options are selected, 30 by a single agent (red) 
and 35 by two agents (blue). Competition forces
the agents to accept inequalities in terms of 
a wide range of different rewards.
{\em Top:} Two example strategies. For most qualities
$q_i$ the payoffs $E^\alpha(q_i)$ (red/blue symbols) 
fall on top of each other.
{\em Bot:} All $M$ strategies. In order to avoid 
overcrowding symbols are not shown.
}
\label{fig_100_100_03_00}
\end{figure}
%%%%%%%%%%%%%%%%%%%%%%%%%%%%%%%%%%

%%%%%%%%%%%%%%%%%%%%%%%%%%%%
\section{Model}
%%%%%%%%%%%%%%%%%%%%%%%%%%%%

The shopping trouble model is defined directly
in terms of strategies, which are given
by the probabilities $p^\alpha(q_i)$ that agent
$\alpha$ selects option $i$. The quality $q_i$
corresponds to the numerical value associated 
with option $i$. For convenience we consider equidistant 
qualities $q_i\in[0,2]$. The support of a strategy 
is given by the set of options for which $p^\alpha(q_i)>0$. 
A strategy is pure when the support contains a 
single option, and mixed otherwise. In the model
$M$ agents have $N$ options to select from, where 
$N$ may be either smaller or larger than $M$.

The payoff an agent receives when selecting $q_i$
is $E_i^\alpha$. On the average, agents receive 
the reward $R^\alpha$,
\begin{equation}
R^\alpha = \sum_i p^\alpha(q_i) E_i^\alpha,
\qquad\quad
\bar{R}=\frac{1}{M}\sum_\alpha R^\alpha\,,
\label{R_alpha}
\end{equation}
where we have defined also the mean reward $\bar{R}$
of all agents. For the shopping trouble model the 
payoff $E_i^\alpha$ contains three terms:
\begin{equation}
E_i^\alpha = v(q_i) -\kappa
\sum_{\beta\ne \alpha} p^\beta(q_i)
+\varepsilon\, p^\alpha(q_i)\log\left(\frac{R^\alpha}{\bar{R}}\right)\,.
\label{ST_model_envy}
\end{equation}
The first term, $v(q_i)=1-(1-q_i)^2$, is the underlying
utility function, compare Fig.~\ref{fig_100_100_03_00}.
Its functional form, as an inverted parabola, is motivated
by the shopping analogy. In this case, products having a bare 
utility $u(q_i)$ can be acquired at a price $q_i$ in the $i$th 
shop. The bare utility should be concave, in view of the law of 
diminishing utility \cite{kauder2015history}, say
$u(q_i) = a\log(q_i+1)$. The utility entering (\ref{ST_model_envy}),
$v(q_i) = u(q_i)-q_i$, is in this case well approximated 
by an inverted parabola.

The second term in (\ref{ST_model_envy}) encodes competition.
A penalty $\kappa(m-1)$ is to be paid by all $m$ agents
when these $m$ agents decide for the same option. It is troublesome,
in the shopping analogy, to buy something in a crowded shop.
Encoding competition directly in terms of the strategy,
as done in (\ref{ST_model_envy}), is an adaptation of the 
framework used commonly for animal conflict models 
\cite{rusch2017logic}, such as the war of attrition
and all pay auctions.

The third term in (\ref{ST_model_envy}) encodes the
desire to compare one's own success, the reward
$R^\alpha$, with that what others receive. As a
yardstick, the average reward $\bar{R}$ has been 
taken, with the envy $\varepsilon$ encoding the 
intensity of the comparison. The log-dependency, 
$\log(R^\alpha/\bar{R})$, is consistent with the 
Weber-Fechner law, namely that the brain discounts 
sensory stimuli, numbers and time logarithmically
\cite{hecht1924visual,dehaene2003neural,howard2018memory}.
Equivalent logarithmic dependencies have been found
for the production of data \cite{gros2012neuropsychological},
and decision-induced chart rankings \cite{schneider2019five}.
For small relative deviations from the mean, when
$\delta R^\alpha =(R^\alpha-\bar{R})/\bar{R}\ll1$, 
the envy term becomes linear,
$\log(R^\alpha/\bar{R})\sim \delta R^\alpha$.
Envy is then directly proportional to 
$R^\alpha-\bar{R}$, a functionality that is
equivalently at the basis of status seeking games 
\cite{haagsma2010equilibrium}. In effect, the
rational behind the envy term is straightforward.
When happy, when $\log(R^\alpha/\bar{R})>0$, the
agent reinforces the current strategy, which is
encoded by $p^\alpha(q_i)$, trying to change it 
instead when $\log(R^\alpha/\bar{R})<0$.

The utility function (\ref{ST_model_envy}) 
of the shopping trouble model can be considered to
encode status seeking, albeit indirectly. Agents
try to maximize utility not only in absolute,
but also in relative terms, with the envy parameter
$\varepsilon$ determining the relative weight of the
two contributions. Outperforming others corresponds
in this interpretation to increased levels of social
status. In difference to standard status seeking 
games \cite{congleton1989efficient}, for which agents 
follow two separate objectives, utility and status 
maximization, the shopping trouble model contains only 
a single, combined utility. The issue of Pareto 
optimality does hence not arise.

The bare formulation of the here introduced shopping trouble 
game, as given by (\ref{ST_model_envy}), is supplemented 
by the concept of migration. One postulates that
agents receiving negative rewards leave the society
in search for better opportunities. Better no
reward at all than to engage with detrimental returns.
Negative rewards appear for large $\kappa$ and
elevated densities $\nu=M/N$ of agents per options,
e.g.\ necessarily when $M=2N$ and $\kappa>1$. For
$M<N$ there are in contrast always Nash equilibria
for which all individual rewards are positive. 
Migration is induced additionally by the envy term, as 
$\log(R^\alpha/\bar{R})$ diverges negatively for
$R^\alpha\to0$. Numerically we solved the shopping 
trouble model using standard replicator 
dynamics \cite{hofbauer2003evolutionary},
\begin{equation}
p_i^\alpha(t+1) = \frac{p_i^\alpha(t)E_i^\alpha(t)}
{\sum_j p_j^\alpha(t)E_j^\alpha(t)}\,.
\label{evolutionaryDynamics}
\end{equation}
For a smooth convergence one adds a constant offset $E_0$ 
to the payoffs on the right-hand side. The offset helps 
in particular to avoid the occurrence of negative rewards, 
which can arise intermediately when a time evolution scheme 
is discrete in time, as for (\ref{evolutionaryDynamics}).
Typically we took $E_0=20$, iterating $5\cdot10^5$ 
times. A defining feature of the shopping trouble model is 
that all agents have functionally identical payoffs.
Only the starting strategies, which we did draw
from a flat distribution, differentiate between 
agents.

%%%%%%%%%%%%%%%%%%%%%%%%%%%%%%%%%%
\begin{figure}[t!]
%\centerline{
%\includegraphics[width=1.00\columnwidth,angle=0]{fig_100_100_03_both.pdf}
%           }
\centerline{
\includegraphics[width=1.00\columnwidth,angle=0]{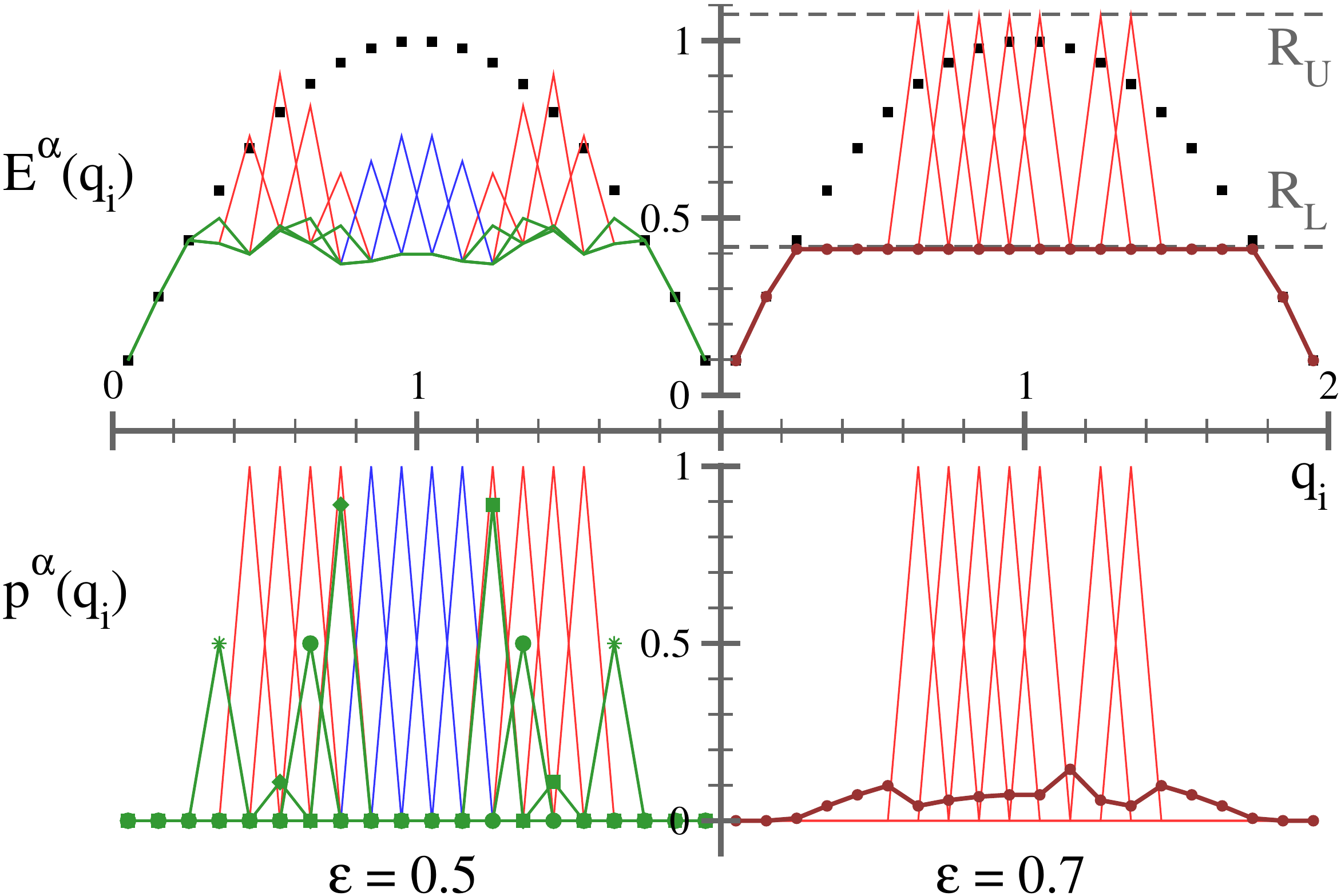}
           }
\caption{{\bf Spontaneous policy reorganization.} 
For $\kappa\!=\!0.3$, $M\!=\!20$ agents and $N\!=\!20$ options,
the payoffs $E^\alpha(q_i)$, as connected by lines (top panels),
and the respective strategies $p^\alpha(q_i)$ (bottom panels).
For $\varepsilon\!=\!0.5$ (left panels) and $\varepsilon=0.7$ (right panels). 
For the simulations (\ref{evolutionaryDynamics}) has been used.
Symbols have been added for clarity to the four individual mixed 
strategies in the lower-left panel. For $\varepsilon\!=\!0.7$ 
seven agents play pure strategies (red), the other 13 agents 
the identical mixed strategy (brown). Included in the upper-right 
panel are the theory predictions for the lower/upper class 
rewards $R_{\rm L}$ and $R_{\rm U}$ (dashed lines), as given by
(\ref{R_L}) and (\ref{R_U}).
Compare Fig.~\ref{fig_100_100_03_00} for $\varepsilon\!=\!0$.
}
\label{fig_100_100_03_both}
\end{figure}
%%%%%%%%%%%%%%%%%%%%%%%%%%%%%%%%%%
%    aLog, kappa, envy    :   2.0000   0.3000   0.5000
%      policyClusterCount :       16
%       nPureStrategies_M :       16
%   nPureSingleStrategies :        8
%   nPureDoubleStrategies :        4
%        nMixedStrategies :        4

%    aLog, kappa, envy    :   2.0000   0.3000   0.7000
%      policyClusterCount :        8
%       nPureStrategies_M :        7
%   nPureSingleStrategies :        7
%   nPureDoubleStrategies :        0
%        nMixedStrategies :        1

%%%%%%%%%%%%%%%%%%%%%%%%%%%%
\section{Results}
%%%%%%%%%%%%%%%%%%%%%%%%%%%%

In the absence of envy, when $\varepsilon=0$, agents
just need to compare the payoff $v(q_i)-\kappa$
of options already taken by somebody else to the
one's that are still available. For $\kappa=0.3$ 
the outcome is presented
in Fig.~\ref{fig_100_100_03_00}. Qualities with 
larger utilities are doubly taken, lower returning
options on the other hand only by a single agent.
The resulting Nash state is unique. Agents avoid
each other, as far as possible, which could be
interpreted as cooperation. Cooperation is 
however not voluntary, but forced by the penalty
$\sim\!\kappa$ incurring when not cooperating.
A consequence of forced cooperation is that the
payoffs received by individual agents vary
considerably. This is notable, as all players
start out equal, differing only with respect to
their initial strategies.

The forced cooperating state is modified once
$\varepsilon$ becomes finite, retaining however
its overall character for moderate envy.
Altogether two types of multi-agent Nash 
equilibria are observed.
\begin{itemize}
\item {\bf Forced Cooperation.} The distribution of
      rewards is continuous. Pure strategies dominate. 
      With increasing envy mixed strategies become 
      more frequent.
      Stable for small to intermediate $\varepsilon$.
\item {\bf Class Separation.}
      The society separates strictly into an upper and a lower class.
      Upper class agents play exclusively pure strategies,
      all lower class agents the identical mixed strategy.
      Agents belonging to the same class receive identical rewards.
      The number of upper class agents decreases monotonically
      with increasing envy, towards one, the monarchy state. 
      Stable for larger $\varepsilon$. 
\end{itemize}
For an initial illustration we concentrate on a small system,
with $M=N=20$, as presented in Fig.~\ref{fig_100_100_03_both}.
One observes that forced cooperation dominates for 
$\varepsilon=0.5$, but with some pronounced differences
to the case $\varepsilon=0$, see Fig.~\ref{fig_100_100_03_00}.
The support of pure and mixed strategies, which develop for 
finite envy, overlap at times, which induces varies 
levels of competition. The supports of different mixed
strategies are distinct.

%%%%%%%%%%%%%%%%%%%%%%%%%%%%%%%%%%
\begin{figure}[t!]
%\centerline{
%\includegraphics[width=1.00\columnwidth,angle=0]{fig_100_100_03_04.pdf}
%           }
%\centerline{
%\includegraphics[width=1.00\columnwidth,angle=0]{fig_100_100_03_08.pdf}
%           }
\centerline{
\includegraphics[width=1.00\columnwidth,angle=0]{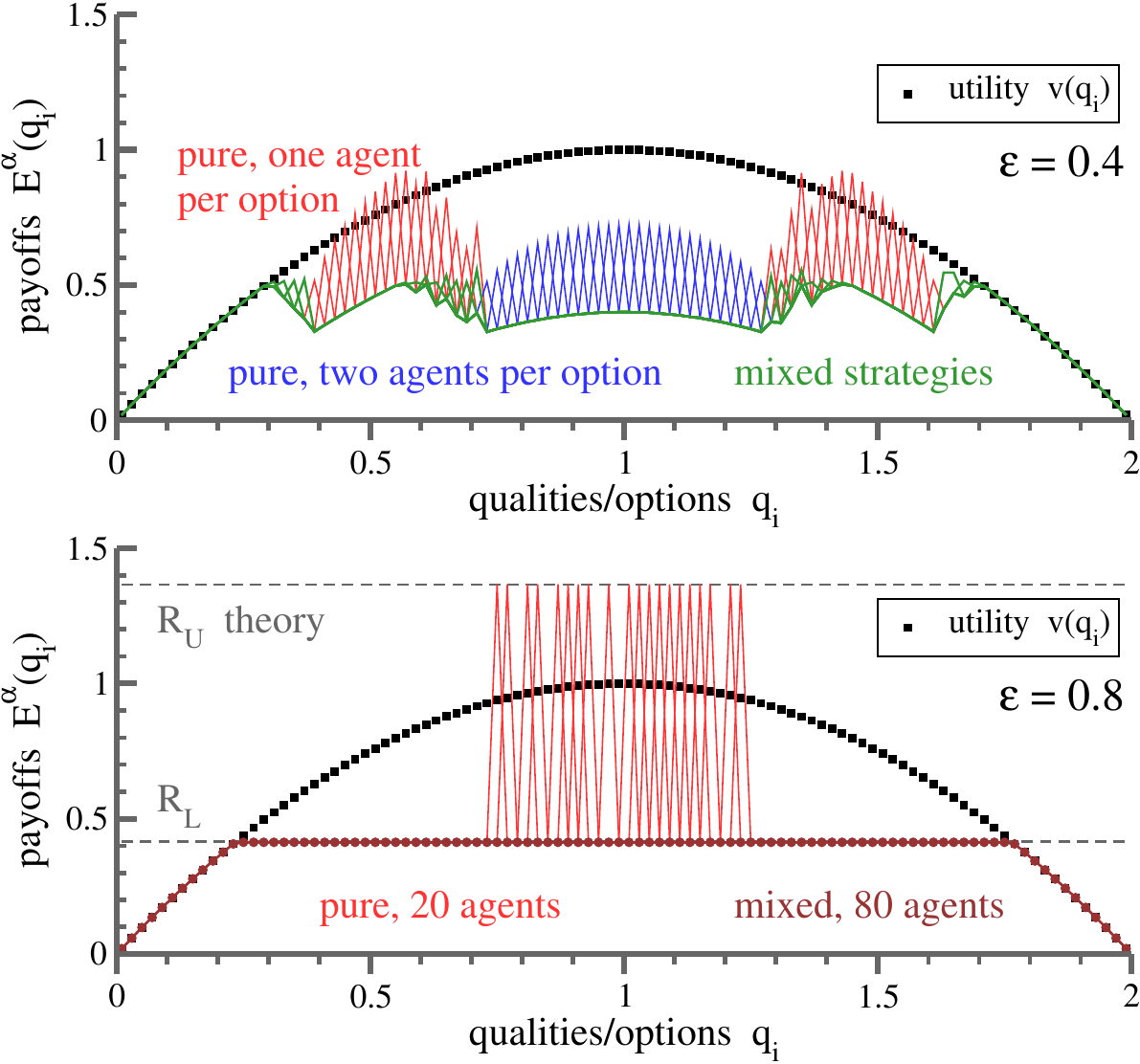}
           }
\caption{{\bf Self-induced class separation.} For $M\!=\!N\!=\!100$ 
and $\kappa\!=\!0.3$, the payoffs $E^\alpha(q_i)$ for the 
two types of Nash equilibria found, forced cooperation,
and class separation (top/bottom).
{\em Top}: For $\varepsilon=0.4$. The number of agents
playing mixed (green) and pure (red/blue) strategies are
10 and 90. A total of 62 pure qualities are selected,
34/28 by one/two agents.
{\em Bottom}: For $\varepsilon=0.8$. Pure strategies (red) are
played by the 20 upper class agents, with all 80 lower class
agents using the identical mixed strategy (brown). Also shown
are the universal theory predictions (\ref{R_L}) and (\ref{R_U})
for the lower/upper class rewards $R_{\rm L}$ and $R_{\rm U}$ 
(dashed lines). The reorganization of the payoff 
spectrum is a collective effect.
}
\label{fig_100_100_03_0408}
\end{figure}
%%%%%%%%%%%%%%%%%%%%%%%%%%%%%%%%%%

% --------- % % --------- % % --------- %
\subsection{Endogenous class stratification}
% --------- % % --------- % % --------- %

A complete self-organized reorganization of the 
spectrum of policies is observed with raising 
strength of envy. The society of agents separates
on its own into two distinct classes, an upper 
and a lower class. The payoff functions of all 
agents are identical, which implies that this
transition, as seen in Fig.~\ref{fig_100_100_03_both}
when going from $\varepsilon=0.5$ to $\varepsilon=0.7$,
is a collective effect. The initial state of the system
determines uniquely where a given agent ends up. The
class-stratified state has several conspicuous 
properties.
\begin{itemize}
\item Upper class agents follow exclusively pure strategies,
      avoiding competition in most cases.
\item A single mixed strategy develops, played by
      the entirety of lower class agents. The support
      of the lower class mixed strategy covers all 
      upper class pure strategies.
\item Only two levels of rewards are present, one
      for each class.
\end{itemize}
That the payoff function $E^\alpha(q_i)$ of the lower class
is  constant on the support of the lower class mixed strategy 
is a necessary condition for an evolutionary stable 
strategy \cite{smith1974theory}.
It would be favorable to readjust the $p^\alpha(q_i)$ if 
this was not the case. It is also not surprising that all 
lower class agents receive the same reward $R_{\rm L}$, 
given that they play identical strategies. Highly non-trivial
is however, that the rewards of all upper class agents
coincide. This can be proven analytically, as
done in the Methods section. For $R_{\rm L}$ the expression
\begin{equation}
R_{\rm L} = \varepsilon\,
\frac{1-f_{\rm L}}{\mathrm{e}^{\kappa/\varepsilon}-1}
\log\left(\frac{\mathrm{e}^{\kappa/\varepsilon}-f_{\rm L}}{1-f_{\rm L}} \right)
\label{R_L}
\end{equation}
is exact when upper class policies are unique, viz if no
option is taken by more than one agent, which hold for most instances. 
The only free parameter in (\ref{R_L}) is the fraction of 
lower class agents, $f_L$, which needs to be determined numerically.
For the Nash state shown in Fig.~\ref{fig_100_100_03_both}
one finds $f_{\rm L}=13/20$ for $\varepsilon=0.7$. The 
resulting prediction (\ref{R_L}) for the reward $R_{\rm L}$ of 
the lower agrees remarkably well with numerics, as seen in
Fig.~\ref{fig_100_100_03_both}. The analytic prediction 
for the reward of the upper class, $R_{\rm U}$, is
\begin{equation}
R_{\rm U} = \varepsilon\,
\frac{1-f_{\rm L}\mathrm{e}^{-\kappa/\varepsilon}}{1-\mathrm{e}^{-\kappa/\varepsilon}}
\log\left(\frac{\mathrm{e}^{\kappa/\varepsilon}-f_{\rm L}}{1-f_{\rm L}} \right)\,,
\label{R_U}
\end{equation}
as derived in the Methods section. Again, theory 
and numerical simulations are in agreement.

The underlying utility $v(q_i)$ enters the theory 
expressions for $R_{\rm L}$ and $R_{\rm U}$ only 
implicitly, via the fraction $f_{\rm L}$ of lower 
class agents, but not explicitly. The properties 
of class-stratified states with the same $\kappa$, 
$\varepsilon$, and $f_L$ are hence identical and 
independent of the shape of the utility function. 
We tested this proposition performing simulations
using a triangular utility function, $v(q_i)=1-|1-q_i|$,
finding that (\ref{R_L}) and (\ref{R_U}) hold perfectly.
Class-stratification leads as a consequence to a Nash state
with universal properties, the telltale sign of a 
collective effect. 

%%%%%%%%%%%%%%%%%%%%%%%%%%%%%%%%%%
\begin{figure}[t]
%\centerline{
%\includegraphics[width=1.0\columnwidth,angle=0]{fig_100_100_03_mixed.pdf}
%           }
\centerline{
\includegraphics[width=1.0\columnwidth,angle=0]{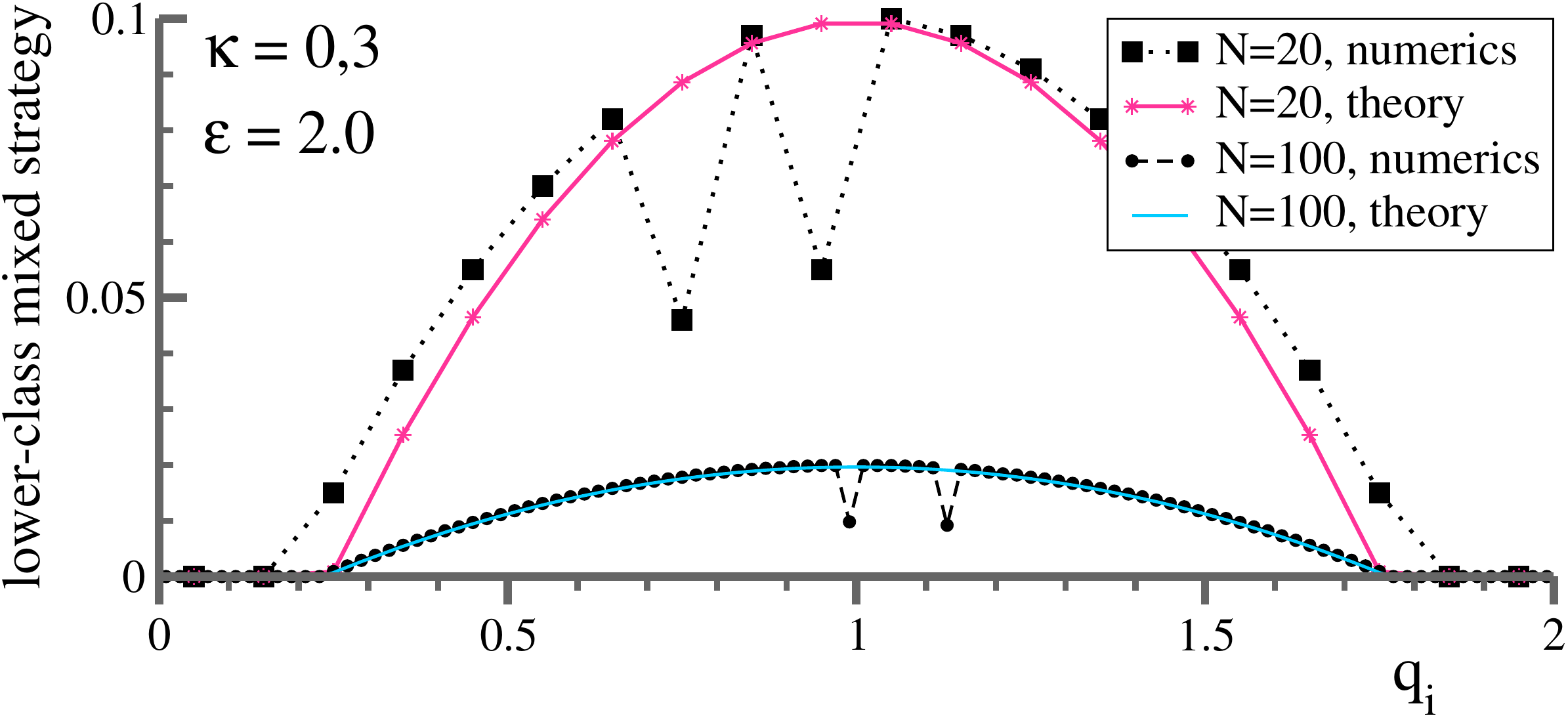}
           }
\caption{{\bf Mixed lower class strategy.} For 
$\kappa\!=\!0.3$ and $\varepsilon\!=\!2.0$, the 
mixed strategy $p^\alpha(q_i)$ identical to all 
lower class agents. Shown are the results of numerical 
simulations (black) and the respective large-$N$ theory 
prediction (\ref{p_mixed_f_L_0}). The dips in the 
numerical result are due to the presence of two upper 
class agents (not shown), respectively for both $M\!=\!N\!=\!20$ 
and $M\!=\!N\!=\!100$. Compare Fig.~\ref{fig_100_100_03_both}.
The improved agreement between simulations and theory 
between $N\!=\!20$ and $N\!=\!100$ is consistent with the 
precondition of the theory, which becomes asymptotically 
exact in the large-$N$ limit.
}
\label{fig_100_100_03_mixed}
\end{figure}
%%%%%%%%%%%%%%%%%%%%%%%%%%%%%%%%%%

That upper class agents receive identical payoffs
is an interesting aspect of universality. It is possible
because the lower class agents adapt their strategies
such that the functional dependence of $v(q_i)$ 
on the qualities is exactly compensated by the
competition term $\sim\!\kappa$. Evidence for this
mechanism can be seen in Fig.~\ref{fig_100_100_03_both}
for $\kappa=0.7$.

The transition from forced cooperation to a stratified society 
is found for all system sizes. In Fig.~\ref{fig_100_100_03_0408} 
we present to this end simulations for $M=N=100$. Below the 
transition, here for $\kappa=0.4$, one observes that individual 
mixed strategies start out at the fringes of the support regions 
of the pure strategies. For $\varepsilon=0.8$ a stratified society 
is present, with the vast majority of agents, 80\%, playing one 
and the same mixed strategy. These agents form the lower class.

% --------- % % --------- % % --------- %
\subsection{Lower class mixed strategy}
% --------- % % --------- % % --------- %

When forced cooperation is present, qualities $q_i$ with high 
utilities $v(q_i)$ are selected without exception by agents
playing pure strategies. This is not the case for the 
spectrum of upper class policies, which may have gaps 
in the class-stratified state, as evident both in
Fig.~\ref{fig_100_100_03_both} and 
Fig.~\ref{fig_100_100_03_0408}. At first sight, this could
seem a contradiction to Nash stability. Lower class
agents are however more likely to visit a quality
$q_i$ not selected by the upper class, as can 
be seen in Fig.~\ref{fig_100_100_03_both}, which
leads to a competition block proportional to the
competition penalty $\kappa$. It is hence not 
favorable for upper class agents to switch. The 
occurrence of gaps implies in particular that the 
Nash state is not unique. 

The lower class mixed strategy has a well defined 
functional form, $p_{\rm mix}(q_i)=p^\alpha(q_i)$,
in the limit $f_U\to0$, namely
\begin{eqnarray}
\label{p_mixed_f_L_0}
p_{\rm mix}(q_i) &=& \frac{1}{\kappa(M-1)}\Big[v(q_i)-E_{\rm c}\Big]
%, \qquad\quad
\\[0.5ex]
E_{\rm c} &=& 1- \left(\frac{3\kappa}{2}\right)^{2/3}
\left(\frac{M-1}{N}\right)^{2/3}\,,
\nonumber
\end{eqnarray}
an expression derived in the Methods section. The mixed
strategy given by (\ref{p_mixed_f_L_0}) is clearly 
non-universal, being linear in the utility 
$v(q_i)$, in contrast to the lower- and upper-class rewards.
The expression for the reward $E_{\rm c}$ is specific
to the utility function $v(q_i) = 1-(1-q_i)^2$, 
and hence equally non-universal.

The simulations presented in Fig.~\ref{fig_100_100_03_mixed}
for $M=N=20$ and $M=N=100$ show that (\ref{p_mixed_f_L_0})
approximates the data well when the system is large. Of
interest are in particular the dips in the data for the
mixed strategy, which occur for qualities $q_i$ selected
by upper class agents. These dips are essential for attaining
universal $R_{\rm L}$ and $R_{\rm U}$, as laid out in 
the Methods section. Note that $p_{\rm mix}(q_i)$ 
is normalized, $\sum_i p_{\rm mix}(q_i)=1$, when the
qualities are dense.

%%%%%%%%%%%%%%%%%%%%%%%%%%%%%%%%%%
\begin{figure}[t]
%\centerline{
%\includegraphics[width=1.0\columnwidth,angle=0]{fig_100_100_03_envy.pdf}
%           }
\centerline{
\includegraphics[width=1.0\columnwidth,angle=0]{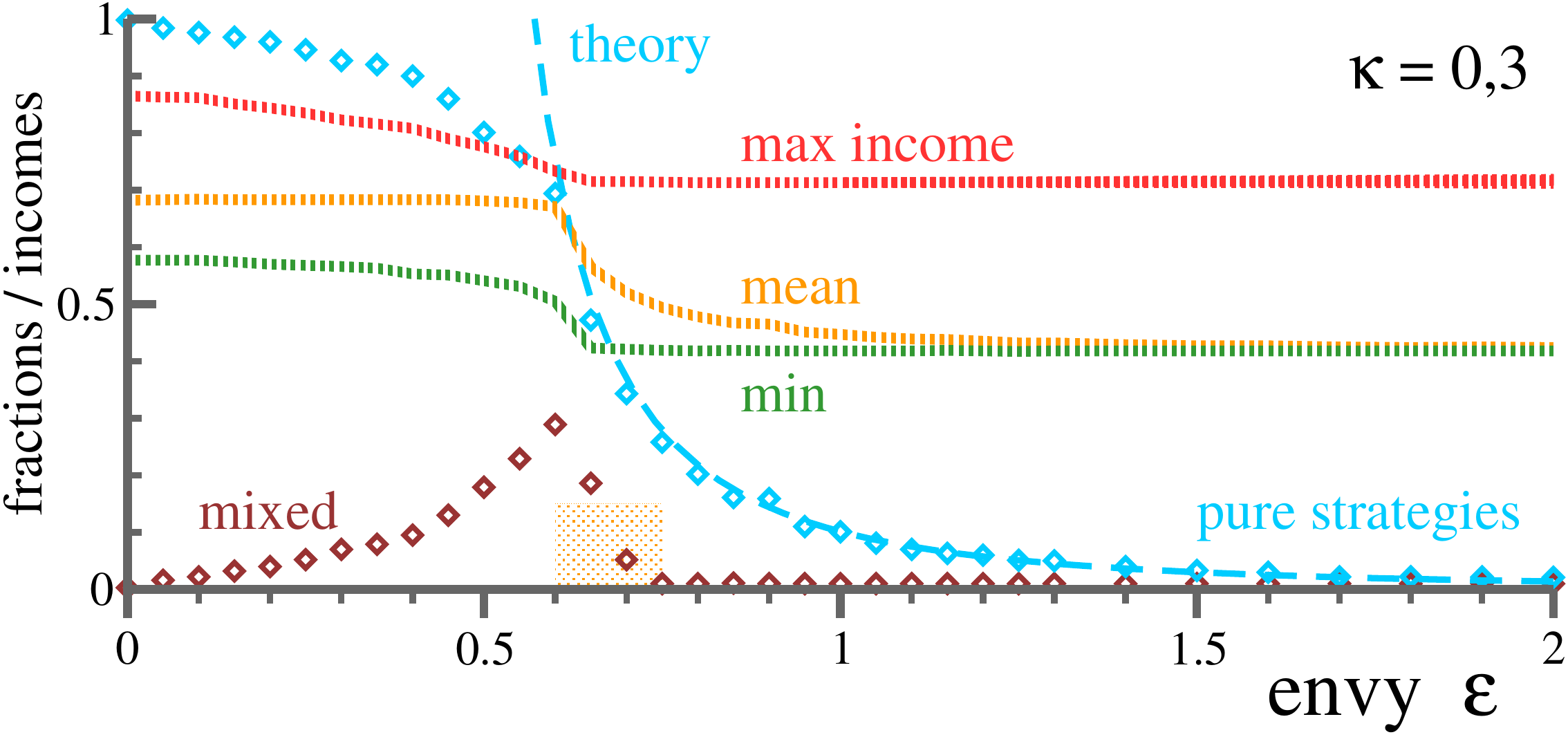}
           }
\caption{{\bf Characteristics of the Nash state as 
a function of envy.} For $\kappa\!=\!0.3$ and 
$M\!=\!N\!=\!100$ the properties of the evolutionary 
stable stationary state. Numerical results are 
averaged over at least ten random initial conditions. 
Shown is the number of distinct mixed strategies per 
agent (brown) and the percentage of agents playing pure 
strategies (blue). Denoted by `theory' is the estimate 
for the fraction of upper class agents, as determined 
by the self-consistency condition (\ref{self_consistency_f_U}), 
which becomes exact for large $\varepsilon$. Also 
included are the maximal (red), the mean (yellow) and
the minimal (green) monetary incomes, compare (\ref{I_alpha}). 
Forced cooperation and class separation are fully present 
respectively for $\varepsilon<0.6$ and 
$\varepsilon>0.75$. In between (shaded orange), both
states may be reached when starting from random initial
policies.
}
\label{fig_100_100_03_envy}
\end{figure}
%%%%%%%%%%%%%%%%%%%%%%%%%%%%%%%%%%

% --------- % % --------- % % --------- %
\subsection{Monarchy vs.\ communism}
% --------- % % --------- % % --------- %

We performed numerical simulations for a wide
range of parameters, mostly for $N=100$ options.
In order to check for finite-size effects we
compared selected parameters settings with
simulations for $N=500$, finding only minor effects.
In Fig.~\ref{fig_100_100_03_envy} representative
data for $\kappa=0.3$ and $M=N=100$ is presented.
Varying the filling fraction $M/N$ and/or $\kappa$
shifts the locus of the transition, leading otherwise
only to quantitative changes. We define with 
$N_{\rm pure}$ the number of agents playing
pure strategies and with $N_{\rm mix}$ the number
of distinct mixed strategies

For small $\varepsilon$ the fraction of mixed 
strategies $N_{\rm mix}/M$ raises monotonically, 
as shown in Fig.~\ref{fig_100_100_03_envy}, 
attaining a maximum when the transition from forced 
cooperation to class stratification starts to take 
place, here at $\varepsilon\approx0.6$. The width of 
the transition is finite, in the sense that either 
state may be reached when starting from random initial 
conditions. As a test we ran twenty independent 
simulations for $\kappa=0.3$ and $\varepsilon=0.65$, 
finding that about half led to forced collaboration 
and half to class separation.  Overall, at least ten 
random initial strategies have been used for the 
individual data points presented in 
Fig.~\ref{fig_100_100_03_envy}. The transition to class 
stratification is completed when the number of mixed 
strategies drops to one, which is the case for 
$\kappa=0.3$ for $\varepsilon\approx0.75$.

The fraction of agents $N_{\rm pure}/M$ playing 
pure strategies decreases monotonically for all 
$\varepsilon$, with the decrease accelerating 
in the transition region from forced collaboration
to class separation. For large values of envy, 
roughly for $\varepsilon>2.2$ a monarchy state 
is reached. The number of upper class agents is
now minimal, mostly just one, occasionally also 
two. An alternative to monarchy 
would be communism, namely that the entire society 
of agents adopts $p_{\rm mix}(q_i)$, as defined 
by (\ref{p_mixed_f_L_0}). We find communism never 
to be stable, both when starting with random initial 
condition and when starting close to the communist 
state. For the latter we performed 
simulations for which the initial strategies
of all agents was $p_{\rm mix}(q_i)$, plus a
perturbation consisting of 1\% relative noise.

Included in Fig.~\ref{fig_100_100_03_envy} 
is an approximate analytic prediction 
for the fraction $f_U=1-f_L$ of upper class agents,
which is obtained from solving
\begin{equation}
1- \left(\frac{3\kappa(M-1)}{2N}\right)^{2/3} =
\varepsilon\,\frac{1-f_{\rm L}}{\mathrm{e}^{\kappa/\varepsilon}-1}
\log\left(\frac{\mathrm{e}^{\kappa/\varepsilon}-f_{\rm L}}{1-f_{\rm L}} 
\right)
\label{self_consistency_f_U}
\end{equation}
self-consistently for $f_L$. The derivation of
(\ref{self_consistency_f_U}), which is valid
in the class-separated state for large $\varepsilon$,
$N$ and $M$, is given in the Methods section. The 
agreement with the numerical results for the fraction 
of agents playing pure strategies, which coincides 
with the fraction of upper class agents, is 
remarkable. Of interest is in particular the 
observation that the theory and numerics stay
close down to the transition to forced cooperation.

%%%%%%%%%%%%%%%%%%%%%%%%%%%%%%%%%%
\begin{figure}[t!]
%\centerline{
%\includegraphics[width=1.00\columnwidth,angle=0]{fig_phaseDiagram.pdf}
%           }
\centerline{
\includegraphics[width=1.00\columnwidth,angle=0]{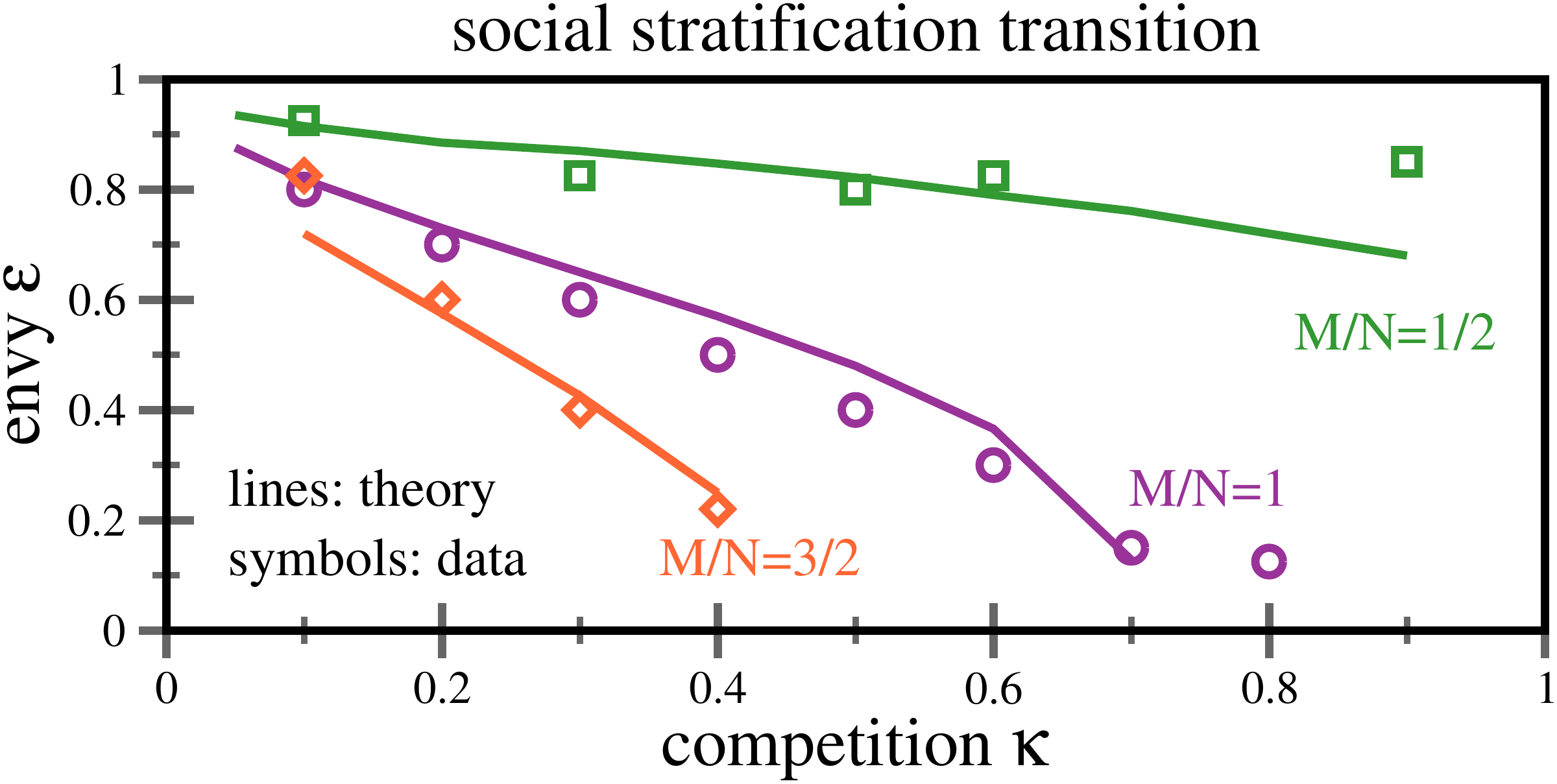}
           }
\caption{{\bf Social stratification transition.}
Phase diagram for the transition from forced cooperation
(below the lines/symbols) to the class separated state
(above the lines/symbols). For the theory the locus 
$f_U\!=\!1/2$, as evaluated by solving (\ref{fig_phaseDiagram}) 
self-consistently, has been used as an indicator.
The data has been obtained for $N\!=\!100$ and respectively
$M\!=\!50/100/150$. In this case the peak in the number of mixed
strategies have been taken as proxies for the transition.
Compare Fig.~\ref{fig_100_100_03_envy}. It is noticeable that 
theory and numerics track each other, in order of magnitude.
The actual width of the transition has not been estimated.
}
\label{fig_phaseDiagram}
\end{figure}
%%%%%%%%%%%%%%%%%%%%%%%%%%%%%%%%%%

% --------- % % --------- % % --------- %
\subsection{Reward vs.\ real-world income}
% --------- % % --------- % % --------- %

The three terms entering the payoff function
(\ref{ST_model_envy}) of the shopping trouble
model are distinct in character. The 
underlying utility $v(q_i)$ and the penalty
arsing from competition, $\sim\!\kappa$,
are real-world monetary payoff terms.
Envy, the propensity to compare one's
own success with that of others, could be 
classified in contrast as being a
predominately psychological component.
Taking this view we define with
\begin{equation}
I^\alpha = R^\alpha - \varepsilon
\log\left( \frac{R^\alpha}{\bar{R}}\right)
\sum_i \big[p^\alpha(q_i)\big]^2
\label{I_alpha}
\end{equation}
the monetary income $I^\alpha$ of agent $\alpha$ as 
the average payoff minus the envy term. 
In the forced cooperation phase the average 
income $\bar{I}$ in nearly constant as a 
function of envy, as shown in 
Fig.~\ref{fig_100_100_03_envy},
dropping however substantially once 
class stratification sets in. In this respect
the society is better off at low to moderated
levels of envy. Class separation does not help
the general public. Also included in
Fig.~\ref{fig_100_100_03_envy} are the
minimal and maximal incomes, 
$I_{\rm min} = \mathrm{min}_\alpha I^\alpha$ and
$I_{\rm max} = \mathrm{max}_\alpha I^\alpha$.
In the class stratified phase $I_{\rm min}$
and $I_{\rm max}$ correspond respectively to the
income of the lower and of the upper class. 
Both are flat, which implies that the drop of 
the mean monetary return $\bar{I}$ with 
increasing envy is due to the simultaneously
occurring decrease in number of upper class agents.

At no stage are incomes increased when envy
is present in a competitive society. This result
holds for the maximal, the minimal and the mean
income, as evident from the data presented
in Fig.~\ref{fig_100_100_03_envy}. It is also
conspicuous that the monetary returns of both the
lower and the upper class are essentially unaffected
by the value of $\varepsilon$, once class separation
sets in. This result is in agreement with
the observation that the lower class mixed strategy
is well approximated by the large $\varepsilon$
limit, as shown in Fig.~\ref{fig_100_100_03_mixed}.
Note that the reward of the upper class diverges
for $\varepsilon\to\infty$, in contrast to the monetary 
income. Of interest is also that $\bar{I}$ remains flat
during forced collaboration, despite the
fact that both $I_{\rm min}$ and $I_{\rm max}$ drop.
This is due to the ongoing reorganization of the
payoff spectrum. 

% --------- % % --------- % % --------- %
\subsection{Phase diagram}
% --------- % % --------- % % --------- %

In Fig.~\ref{fig_phaseDiagram} the phase diagram 
as a function of $\kappa$ and $\varepsilon$,
competition and envy, is presented. Systems with 
$N=100$ options and $M=50/100/150$ agents have been 
simulated numerically. For the onset of the transition 
from forced cooperation to class separation the maximum 
of the number of mixed strategies has been taken as an 
indicator. The level of envy needed for the society
to phase separate decreases with increasing
competition, a somewhat intuitive result.
The same trend holds when increasing the
density $M/N$ of agents per available options,
which makes it more difficult to avoid 
each other.

Included in Fig.~\ref{fig_phaseDiagram} 
are estimates obtained by solving the
self-consistency condition (\ref{self_consistency_f_U})
for $f_U=1/2$. This rough estimate for the 
transition to class stratification tracts the 
numerical results surprisingly well. Deviations 
are seen in particular for larger $\kappa$.

%%%%%%%%%%%%%%%%%%%%%%%%%%%%
\section{Methods}
%%%%%%%%%%%%%%%%%%%%%%%%%%%%

The penalty term  of the shopping trouble model 
(\ref{ST_model_envy}) can be written as
\begin{equation}
\sum_{\beta\ne \alpha} p^\beta(q_i)=
M\bar{p}(q_i)-p^\alpha(q_i),
\quad\quad \bar{p}(q_i) = \frac{1}{M} 
\sum_{\beta} p^\beta(q_i)\,,
\label{ST_penalty_alpha}
\end{equation}
which demonstrates that agents interact via a 
quality-dependent mean field, the average strategy $\bar{p}(q_i) $. 
The number of terms is $N-1$, which could give the impression 
that the shopping trouble model is not size consistent. 
This is however not the case, as both pure and mixed
strategies contribute on the average to the order $1/N$ 
to the sum of the $\kappa$-term. For a fixed occupation density 
$\nu=M/N$, the thermodynamic limit $M,N\to\infty$ is therefore
well defined. Numerically we find that the properties
of the Nash equilibria change only in minor ways when 
increasing $N$, but retaining $\kappa$,
$\varepsilon$ and $\nu$.

% --------- % % --------- % % --------- %
\subsection{Theory for the class separated state}
% --------- % % --------- % % --------- %

In the shopping trouble model agents have 
functionally identical payoff functions, which
implies that a priori distinctions between agents 
are not present. Where an agent ends up, in the lower 
or in the upper class, depends as a consequence solely 
on the respective initial conditions. For our analysis
of the class separated state we denote with $q_{\rm U}$ 
and $q_{\neg\rm U}$ qualities within the support 
of the lower class that are taken/not taken by
upper class agents. $M_L/M_U$ are respectively 
the number of lower/upper class agents. We assume 
that upper class agents are forced to cooperate 
fully, playing distinct options. Numerically this 
holds in most cases.

Central to our consideration are the properties of 
evolutionary stable strategies \cite{smith1974theory},
in particular that the payoff is constant within the 
support. For the mixed strategy of the lower class, 
$p_{\rm L}(q_i)$, this implies that the payoff 
function $E(q_i)$ is identical to the reward $R_{\rm L}$,
\begin{equation}
R_{\rm L} = 
v(q_{\neg\rm U})-\kappa\,[M_{\rm L}-1]\,p_{\rm L}(q_{\neg\rm U}) 
+\varepsilon\,p_{\rm L}(q_{\neg\rm U})\,\log(R_{\rm L}/\bar{R})\,.
\label{condition_q_L_of_LC}
\end{equation}
Playing against an evolutionary stable strategy 
entails to receive the same constant payoff 
\cite{smith1974theory}. Outside the support of their
own pure strategies, upper class agents play against
the lower class and against other upper class agents.
For $q_i$ for which $p_{\rm L}(q_i)>0$ and and for
which all upper class policies vanish, $p_{\rm U}(q_i)=0$, 
the consequence is that $E(q_i)=R_{\rm L}$ also
for upper class agents,
\begin{equation}
R_{\rm L} = \nonumber
v(q_{\neg\rm U})-\kappa M_{\rm L}\, p_{\rm L}(q_{\neg\rm U})\,.
\label{condition_q_L_of_UC}
\end{equation}
Numerically small deviations from (\ref{condition_q_L_of_UC}) 
can occur. The reason is that (\ref{condition_q_L_of_UC}) enters
the evolution equation (\ref{evolutionaryDynamics}) 
multiplied by $p_U(q_{\neg\rm U})$, which vanishes 
in the final state. The condition, that playing against 
an evolutionary stable strategy must yield the payoff 
of the strategy in question, can be enforced consequently
only while $p_U(q_{\neg\rm U})$ is still finite. This is
the case only during a transition period, while iterating 
towards stationarity.

Equating (\ref{condition_q_L_of_LC}) and 
(\ref{condition_q_L_of_UC}) and dividing by 
$p_{\rm L}(q_{\neg\rm U})$, which is positive
within the support of the lower class, yields
the universal relation
\begin{equation}
\log\left(\frac{R_{\rm L}}{\bar{R}}\right) =-\frac{\kappa}{\varepsilon},
\qquad\quad
R_{\rm L}=\bar{R}\,\mathrm{e}^{-\kappa/\varepsilon}\,,
\label{solution_R_low}
\end{equation}
which we verified numerically. Note that 
(\ref{solution_R_low}) is independent of the 
underlying utility function $v(q_i)$, of $N$
and of $M$. Denoting with $f_{\rm U}=M_{\rm U}/M$ and 
$f_{\rm L}=M_{\rm L}/M$ the relative fractions
of upper and lower class agents, one has 
\begin{equation}
\bar{R} = f_{\rm U} R_{\rm U} + f_{\rm L} R_{\rm L},
\qquad\quad
R_{\rm U} = \frac{\bar{R}}{f_{\rm U}}
\left(1-f_{\rm L}\mathrm{e}^{-\kappa/\varepsilon}\right)
\label{solution_R_upper}
\end{equation}
for $R_{\rm U}$, when using (\ref{solution_R_low}) 
for $R_{\rm L}$. For (\ref{solution_R_upper}) we assumed 
that the rewards $R^\alpha$ for upper class agents are
all identical, which we will prove shortly. Together, 
one finds
\begin{equation}
R_{\rm U}-R_{\rm L} = \frac{\bar{R}}{f_{\rm U}}
\left(1-\mathrm{e}^{-\kappa/\varepsilon}\right)
\label{solution_R_gap}
\end{equation}
for the gap in the rewards received by the upper and the lower
class.

Eqs.~(\ref{condition_q_L_of_LC}) and 
(\ref{condition_q_L_of_UC}) are conditions 
for the $q_{\neg\rm U}$, that is for options
not taken by upper class agents. When playing 
an option $q_{\rm U}$ occupied by an upper class 
agent, the lower class payoff function reads
\begin{eqnarray}
\nonumber
R_{\rm L} &=&
v(q_{\rm U})-\kappa\big[(M_{\rm L}-1)\, p_{\rm L}(q_{\rm U})+1\big]
%\\ \nonumber &&\quad 
+\,\varepsilon\,p_{\rm L}(q_{\rm U})\,\log(R_{\rm L}/\bar{R})
\\ &=& 
v(q_{\rm U})-\kappa\,M_{\rm L}\, p_{\rm L}(q_{\rm U})-\kappa\,,
\label{condition_q_U_of_LC}
\end{eqnarray}
when using (\ref{solution_R_low}) and the precondition 
that there is exactly one upper class agent with 
$p_{\rm U}(q_{\rm U})=1$. Note that 
payoff and reward coincide for lower class agents. We now turn 
to the payoff of upper class agents,
\begin{equation}
E_{\rm U}^\alpha = v(q_{\rm U})-\kappa\,M_{\rm L}\,p_{\rm L}(q_{\rm U}) 
+\varepsilon\,\log(R_{\rm U}^\alpha/\bar{R})\,.
\label{condition_q_U_of_UC}
\end{equation}
Here $p_{\rm U}(q_{\rm U})=1$ has been used.
With (\ref{condition_q_U_of_LC}) one obtains
\begin{equation}
E_{\rm U}^\alpha-R_{\rm L} = \kappa +\varepsilon\,
\log(R_{\rm U}^\alpha/\bar{R}), \qquad\quad
E_{\rm U}^\alpha \to R_{\rm U} \,,
\label{R_U_L}
\end{equation}
which is manifestly independent of $q_{\rm U}$, and
hence also of the bare utility function $v(q_i)$. The
independency of (\ref{R_U_L}) with respect to the utility 
function implies that the payoffs of all upper class agents 
coincide, namely that $E_{\rm U}^\alpha \equiv R_{\rm U}$.
Equating (\ref{solution_R_gap}) with (\ref{R_U_L}) 
yields
\begin{equation}
R_{\rm U}-R_{\rm L} = \frac{\bar{R}}{f_U}
\left(1-\mathrm{e}^{-\kappa/\varepsilon}\right)
\,=\, \kappa +\varepsilon\,
\log\left(\frac{R_{\rm U}}{\bar{R}}\right)\,,
\label{gap_1}
\end{equation}
and hence
\begin{eqnarray}
\nonumber
\frac{\bar{R}}{f_U}
\left(1-\mathrm{e}^{-\kappa/\varepsilon}\right)
&=& \kappa +\varepsilon\,
\log\left(\frac{1-f_{\rm L}\mbox{e}^{-\kappa/\varepsilon}}{f_{\rm U}}\right)
\\ &=& \varepsilon\,
\log\left(\frac{\mbox{e}^{\kappa/\varepsilon}-f_{\rm L}}{f_{\rm U}}\right)\,,
\label{gap_2}
\end{eqnarray}
when using (\ref{solution_R_upper}) to eliminate 
$R_{\rm U}/\bar{R}$ on the right-hand-side of (\ref{gap_1}). 
With
\begin{equation}
\bar{R} = \varepsilon\,
\frac{1-f_{\rm L}}{1-\mathrm{e}^{-\kappa/\varepsilon}}
\log\left(\frac{\mathrm{e}^{\kappa/\varepsilon}-f_{\rm L}}{1-f_{\rm L}} \right)
\label{solution_R_mean}
\end{equation}
we obtain a universal relation for the mean reward
$\bar{R}$. It follows, as the argument of 
the logarithm is larger than unity, that $\bar{R}$ is 
strictly positive. The mean reward depends only
implicitly on the utility function $v(q_i)$, through
the fraction $f_{\rm L}$ of lower class agents, but not 
explicitly. Together with (\ref{R_U_L}) and
(\ref{solution_R_gap}) the lower- and upper class 
rewards $R_{\rm L}$ and $R_{\rm U}$ are determined 
as (\ref{R_L}) and (\ref{R_U}).

% --------- % % --------- % % --------- %
\subsection{Identical strategies}
% --------- % % --------- % % --------- %

For the case that all $M$ agents play the
identical strategy $p(q_i)\equiv p^\alpha(q_i)$,
the expected payoff $E_i\equiv E_i^\alpha$ is
\begin{equation}
E_i = v(q_i)-\kappa (M-1) p(q_i) \quad\to\quad E_{\rm c}\,.
\label{ST_E_i}
\end{equation}
With $p(q_i)$ being evolutionary stable, the payoff $E_i$
is constant on the support, $E_i\equiv E_{\rm c}$. For
qualities outside the support, the payoff $E_i$ will 
be lower \cite{smith1974theory}. The probability 
$p(q_i)$ to select an option enters $E_i$ explicitly, 
which implies that $p(q_i)$ is obtained by a 
direct inversion of (\ref{ST_E_i}). One has therefore
\begin{equation}
p(q_i) = \frac{1}{\kappa(M-1)}\Big[v(q_i)-E_{\rm c}\Big]\,.
\label{ST_p_q_i}
\end{equation}
The maxima of the probability distribution $p(q_i)$
and of the utility function $v(q_i)$ coincide. The 
final payoff $E_{\rm c}$ is a free parameter which
is determined by the normalization condition
\begin{equation}
1 = \sum_{i;\,p(q_i)>0} p(q_i),
\qquad\quad p(q_i)\to p_i(E_{\rm c})\,,
\label{ST_normalization_condition}
\end{equation}
where the sum runs over the support of the policy.
For finite $N$ the normalization condition
(\ref{ST_normalization_condition}) needs to
be solved numerically via (\ref{evolutionaryDynamics}).
Results are shown in Fig.\,\ref{fig_100_100_03_mixed}.

% --------- % % --------- % % --------- %
\subsection{Large numbers of options}
% --------- % % --------- % % --------- %

The normalization condition (\ref{ST_normalization_condition})
reduces to an integral for large numbers of qualities $N$.
The boundary of the support are determined by
\begin{equation}
v(q)=E_{\rm c}\, \qquad\quad
q_\pm=1\pm\sqrt{1-E_{\rm c}}\,,
\label{ST_identical_support}
\end{equation}
since $v(q)=1-(1-q)^2$. The normalization 
condition (\ref{ST_normalization_condition})
takes then the form
\begin{equation}
\kappa(M-1) = \int_{-\sqrt{1-E_{\rm c}}}^{\sqrt{1-E_{\rm c}}}
\big(1-x^2-E_{\rm c}\big) \frac{dx}{\Delta x}\,,
\label{ST_normalization_integral}
\end{equation}
when using $x=1-q$ and $\Delta x = 2/N$. We obtain
\begin{equation}
2\kappa\,\frac{M-1}{N} = \left(2-\frac{2}{3}\right) (1-E_{\rm c})^{3/2}\,,
\label{ST_identical_sol}
\end{equation}
which yields (\ref{p_mixed_f_L_0}), or
\begin{equation}
1-E_{\rm c} = \left(\frac{3\kappa}{2}\right)^{2/3}
\left(\frac{M-1}{N}\right)^{2/3}\,.
\label{ST_identical_E_c}
\end{equation}
The resulting mixed strategy $p(q_i)$, as given
(\ref{ST_p_q_i}), is in excellent agreement with
simulations when only a few upper class agents are
left, as illustrated in Fig.~\ref{fig_100_100_03_mixed} 
for $\kappa=0.3$ and $\varepsilon=2$.

Migration occurs when $E_{\rm c}\to0$, that is when
\begin{equation}
1 = \frac{3\kappa}{2}\, \frac{M-1}{N} \approx \frac{3\kappa\nu}{2},
\qquad\quad
\nu =\frac{M}{N}\,,
\label{migration limit}
\end{equation}
where the last approximation holds for large $M$ and $N$. 
The carrying capacity of the society, the maximal possible 
density $\nu$ of agents, scales hence inversely with the 
strength of the competition, $\kappa$. It is independent
on $\varepsilon$.

The fraction $f_U$ of upper class agents is small when
envy is large. In this limit one can approximate the 
lower class reward $R_L$ with $E_{\rm c}$, as determined 
by (\ref{ST_identical_E_c}). This approximation,
$E_{\rm c}\approx R_{\rm L}$, leads 
to (\ref{self_consistency_f_U}), when taking also
(\ref{R_L}) for $R_{\rm L}$ into account.

% --------- % % --------- % % --------- %
\subsection{Scaling for two agents}
% --------- % % --------- % % --------- %

The result for identical strategies, Eq.~(\ref{ST_identical_E_c}),
has a well defined large-$N$ limit for a constant filling
fraction $\nu=M/N$. It is also of interest to consider
the case of finite numbers of agents, $M$, say
$M=2$. In the limit $N\to\infty$ the policy $p(q)$ 
converges in this case towards a pure strategy, viz 
to a delta-function.
The scaling for the maximum $p_{\rm max}$ and the
width $\Delta q$ are
\begin{equation}
p_{\rm max} \sim 1-E_{\rm c} \sim 
\left(\frac{1}{N}\right)^{2/3},
\quad\quad
\Delta q \sim \sqrt{1-E_{\rm c}}
\sim 
\left(\frac{1}{N}\right)^{1/3}\,,
\label{largeNscaling}
\end{equation}
see (\ref{ST_p_q_i}) and (\ref{ST_identical_support}).
The width $\Delta q$ of the support shrinks only slowly 
when increasing the number $N$ of options, remaining 
substantial even for large numbers, such as $N=10^3$. This
is a quite non-trivial result, as one may have expected
that the effect of competition between agents decreases
faster, namely as $1/N$. Note that the scaling of the area, 
$p_{\rm max}\Delta q\sim 1/N$, is determined by the 
density of options, which is $N/2$.

% --------- % % --------- % % --------- %
\subsection{Terminology\label{sect_termiology}}
% --------- % % --------- % % --------- %

For convenience we present here an overview of
the terminology used, including for completeness 
selected key game-theoretical definitions. It follows
that the shopping trouble model is a probabilistic 
competitive evolutionary game based on undifferentiated
but distinguishable agents.

\smallskip\noindent{\bf Options/qualities.}\ 
An option is a possible course of action, like
going to a shop to buy something. For a game with
a large number of options, as considered here, it 
is convenient to associate a numerical value to 
an option. One may either identify the option 
with its numerical value, as it is usual, e.g., for 
the war of attrition, or distinguish them on a
formal level, as done here. For an option $i$ we 
denote with the quality $q_i$ the associated numerical 
value.

\smallskip\noindent{\bf Pure/mixed strategies.}\ 
In simple games, like the Hawk and Dove competition,
options and strategies are often not distinguished. 
Selecting an option, to fight or not to fight, is then
identical to the strategy. On a general level, strategies 
define how and when a player selects one of the possible 
options. A strategy is pure when the agent plays the 
identical option at all times, and mixed otherwise.

\smallskip\noindent{\bf Probabilistic game.}\ 
For probabilistic games strategies are defined
in terms of probabilities. This is the case for
the shopping trouble model, where $p^\alpha(q_i)$
defines the probability that agent $\alpha$ selects
at any time the quality $q_i$ associated with
the option $i$.

\smallskip\noindent{\bf Support.}\ 
A probabilistic strategy assigns a probability
$p^\alpha(q_i)\ge0$ to all possible options. 
One often finds that the $p^\alpha(q_j)$ 
are finite only for a subset of options, the
support of the strategy. The size of the support
is larger than one for mixed strategies, and
exactly one for pure strategies.

\smallskip\noindent{\bf Undifferentiated distinguishable agents.}\ 
Agents are differentiated when every agent is characterized by
an individual set of parameters, and undifferentiated when the
same set of parameters applies to everybody. Strategies are specific 
to individual agents, in any case, when they are distinguishable. 
Indistinguishable agents share in contrast strategies.

\smallskip\noindent{\bf Payoff/reward.}\ 
The payoff function is a real-valued function of the 
qualities/options. The aim is to optimize the
strategy such that the average payoff is maximized. 
For the average payoff the term reward is used 
throughout this study.

\smallskip\noindent{\bf Evolutionary game.}\ 
Evolutionary games are played not just once, but over 
and over again. After each turn, agents update their 
individual strategies according to the payoffs received 
when selecting option $i$ with the probability
$p^\alpha(q_i)$.

\smallskip\noindent{\bf Competitive/cooperative game.}\ 
In cooperative games parties may coordinate their individual
strategies, e.g.\ in order to optimize collective payoffs.
Contracts (like I select option A if you go for B) are,
on the other hand, not possible for competitive games. Also 
possible are coalition formation or hedonic games focusing
on the formation of subgroups.

\smallskip\noindent{\bf Nash equilibrium.}\ 
For competitive games an equilibrium in terms 
of the individual strategies may be attained.
In this state, the Nash equilibrium, rewards 
diminish when individual players attempt to change 
their strategies. More than one Nash state
can exist for identical parameter settings. Nash 
stable configurations of strategies correspond to
locally stable fixpoints of the replicator 
dynamics (\ref{evolutionaryDynamics})
for evolutionary games.

\smallskip\noindent{\bf Collective effect.}\ 
In complex systems theory, a collective effect
is present when the interaction of an extended number
of constituent elements gives rise to a new type
of state. An example from psychology is the emergence
of mass psychology from individual behaviors. In
the shopping trouble model the transition from
forced cooperation to class stratification is
a collective phenomenon.

\smallskip\noindent{\bf Forced cooperation.}\ 
Agents may agree to select different options
in cooperative games, for example in order to
optimize overall welfare. Players may, on the other
hand, be forced to avoid each other in competitive
games, because of the penalties that would
incur otherwise on individual levels. To an outside
observer the resulting state has the traits of
cooperation, which is in this case however 
autonomously enforced.

\smallskip\noindent{\bf Envy.}\ 
In the context of the present study, envy is defined
in terms of the payoff function. For this, the
payoff a given agent $\alpha$ receives, when selecting
a certain option $i$, depends expressively on the rewards 
of the other agents. Envy adds a non-monetary contribution
to the reward of the player, which is positive/negative if
the overall reward of the player is larger/lower
than that of others.

%%%%%%%%%%%%%%%%%%%%%%%%%%%%
\section{Discussion and conclusion}
%%%%%%%%%%%%%%%%%%%%%%%%%%%%

The process of class separation occurring in the 
shopping trouble model has several characteristic 
features. One is that upper- and lower class engage 
in qualitatively different strategies. There are as 
many different pure strategies as there are upper 
class agents, one for each, but only one mixed strategy 
for the entire lower class. Individualism is lost 
when becoming a member of the masses, to put it 
colloquially. Alternatively one may view the common
mixed strategy played by the lower class as
an atypical group-level trait, namely one that
doesn't come with an improved Darwinian fitness 
\cite{smaldino2014cultural}.
For an understanding we note that envy 
enters the shopping trouble model as 
$\varepsilon p^\alpha(q_i)\log(R^\alpha/\bar{R})$,
which implies that the current probability 
$p^\alpha(q_i)$ to select a given quality tends 
to be suppressed when $R^\alpha<\bar{R}$. Envy
has a self-reinforcing effect when the individual
reward $R^\alpha$ is in contrast not smaller, but 
larger than the population average. This argument 
explains why agents with modest/high 
rewards play mixed/pure strategies. 

Evolutionary stable strategies can have different 
rewards only when their supports are not identical, 
which becomes increasingly difficult with the continuous
increase in the number of mixed strategies that is 
observed during forced collaboration with 
raising levels of envy, see Fig.~\ref{fig_100_100_03_envy}. 
Policies merge once the phase space for the support 
of distinct mixed strategies runs out and a single mixed 
lower-class strategy remains. Class stratification 
corresponds from this perspective to a strategy
merging transition, producing in consequence an
atypical group-level trait.

A second feature characterizing class stratification
is universality, namely that the underlying utility
function $v(q_i)$ affects the Nash equilibrium exclusively 
through the fraction $f_{\rm L}$ of lower-class agents.
An interesting corollary is that is does not really matter
which options the upper class selects, as the reward,
and consequently also the monetary income, remains 
unaffected. One could call this freedom the luxury 
of choice of being rich. Upper class strategies tend 
to cluster nevertheless around the maximum of the underlying
utility function, compare Fig.~\ref{fig_100_100_03_0408},
which is however a purely dynamic effect. Policies
that prefer qualities with large $v(q_i)$ have increased 
growth rates while iterating towards stationarity.

Beyond its original interpretation as a competitive
shopping model, one can view the shopping trouble model 
as a basic model for competition for scarce goods,
in particular in a social context. The qualities $q_i$ 
would correspond in this setting either to distinct social 
positions or to job opportunities, with the bare utility 
$v(q_i)$ encoding respectively social status and salaries. 
It is presently unclear to which extent, and if at all, 
human societies can be described in a first approximation 
by the shopping trouble model. In case, western 
societies are presumable in the phase denoted here
forced collaboration, with varying distances to
the class stratification transition. A transition 
to the stratified phase would be equivalent to a 
major socio-cultural paradigm shift \cite{pascual2020epistasis},
such as the possible incipient dynamic instability of
modern democracies due to growing mismatch between 
the build-in time delays, the election cycle, and the
accelerating pace of political opinion dynamics \cite{gros2017entrenched}. 
This is a somewhat worrisome outlook, given that the repercussions of
envy are amplified, as shown in Fig.~\ref{fig_phaseDiagram},
when societies become more and more competitive. A possible 
ongoing development \cite{davies2016limits}.

The stratified phase found in the shopping trouble 
model is the result of a self-organizing process, 
with the consequence that it has universal properties 
that can be controlled only indirectly by external 
influences. Policy makers loose part of their tools 
when a society class separates. Class stratified 
societies are in this sense intrinsically resistant 
to external influences. Overall our results show 
that envy tends to cement class differences, instead 
of softening them. It may be tempting for people at 
the bottom to compare what they have with the 
riches of the top, but it is actually 
counterproductive.

%------------------------------------------------
%------------------------------------------------

\vskip1pc
\ethics{Not applicable.}

\dataccess{This article has no additional data.}

\aucontribute{All work and ideas by C.~Gros.}

\competing{Not applicable.}

\ack{The author thanks Daniel Gros for discussions and
Roser Valenti for valuable suggestions regarding the 
manuscript.}

\funding{Not applicable.}

\disclaimer{Not applicable.}

%------------------------------------------------
%------------------------------------------------

%\vfill
%\pagebreak
%%%%%%%%%%%%%%%%%%%%%%%%%%%%
%\bibliographystyle{unsrt}
%\bibliographystyle{RS}
%\bibliography{royalSociety.bib}
%%%%%%%%%%%%%%%%%%%%%%%%%%%%

%%%%%%%%%%%%%%%%%%%%%%%%%%%%
\end{document}